\documentclass[a4paper,hyphens,keeplastbox]{jacow}
\usepackage[utf8]{inputenc}
\usepackage[english]{babel}
\usepackage[breaklinks=true,hidelinks]{hyperref}

\title{The Factorization Bias in the Van der Meer Method:\break
       \Run2 Experiences at the CMS experiment}
\author{J. Knolle\thanks{\href{mailto:joscha.knolle@cern.ch}{joscha.knolle@cern.ch}}, Deutsches Elektronen-Synchrotron, Hamburg, Germany \\
		on behalf of the CMS Collaboration}

\newcommand{\Nvtx}{N_{\text{vtx}}}
\newcommand{\Mx}{\mathcal{M}_x}
\newcommand{\Ivdm}{\mathcal{I}_{\text{VdM}}}
\newcommand{\Itrue}{\mathcal{I}_{\text{true}}}
\newcommand{\mum}{\textmu m}
\newcommand{\sqrts}[2][]{\smash{$\sqrt{s}_{#1}=#2$\,TeV}}
\newcommand{\delx}{{\Delta x}}
\newcommand{\convV}{\otimes V}
\newcommand{\corr}{\varrho}
\newcommand{\Run}[1]{\mbox{Run #1}}
\newcommand{\RooFit}{\texttt{RooFit}}

\newlength\imagesize
\begin{document}

\maketitle

\begin{abstract}
The luminosity measurement of the CMS experiment at the CERN LHC is calibrated with Van der Meer (VdM) scans. A bias occurs in the VdM method due to the assumption of transversely factorizable proton densities of the LHC beams. Here, the different methods applied in \Run2 to estimate the size of the factorization bias are reported. The beam-imaging method reconstructs the transverse proton densities from beam-imaging scans. Additional methods exploit offset scans, analyze the evolution of the measured luminous region, and evaluate diagonal scans.
\end{abstract}

\section{Introduction}

The CMS experiment~\cite{CMS-Experiment} at the CERN LHC makes use of the Van der Meer (VdM) method to calibrate its luminosity measurement~\cite{LumiDaysMoritz}. Separately for each data-taking period, a set of dedicated VdM scans is performed during an LHC fill with special beam optics. In a VdM scan, the two LHC beams are separated in one transverse direction and then moved in (typically 25) steps across each other. From the event rates measured at each step, the width of the beam overlap region along the scan direction is measured. The product of two widths obtained from orthogonal VdM scans estimates the beam overlap integral, which is used to infer the absolute luminosity scale.

The VdM method assumes that the transverse profile of the beam overlap region factorizes into two one-dimensional profiles along the scan directions. If this assumption does not hold, the VdM estimate of the beam overlap integral is biased. The bias can result from a mismatch between the scan directions and the factorization axes, or (additionally) from an inherent nonfactorization of the transverse profile.

In \Run1 (2009--2013), all LHC experiments observed nonfactorization of the beam overlap region~\cite{LUM-13-001,LhcbLum8Tev,AtlasLum8Tev,AliceLum8Tev}.

In \Run2 (2015--2018), the CMS experiment relied mainly on the beam-imaging method to evaluate the size of the factorization bias. Additionally, new analyses of offset scans, of the evolution of the luminous region (referred to as the ``beamspot''), and of diagonal scans are being developed as complementary methods. This contribution presents these four methods along with their preliminary results.

\section{Beam-Imaging Method}

The beam-imaging method~\cite{BeamImaging}, developed at the CMS experiment, employs a set of four special beam-imaging scans to reconstruct the transverse proton densities of the two LHC beams. These proton densities can then be used to estimate the factorization bias of the VdM method.

\subsection{Beam-Imaging Scans}

A beam-imaging scan is a variant of a VdM scan where one beam is kept at a central transverse position and the other beam is moved in (typically 19) steps across the non-moving beam along one transverse direction. Four beam-imaging scans, for both beams in two orthogonal transverse directions, form a scan set.

\begin{figure}[!h]
\centering
\includegraphics[width=\imagesize]{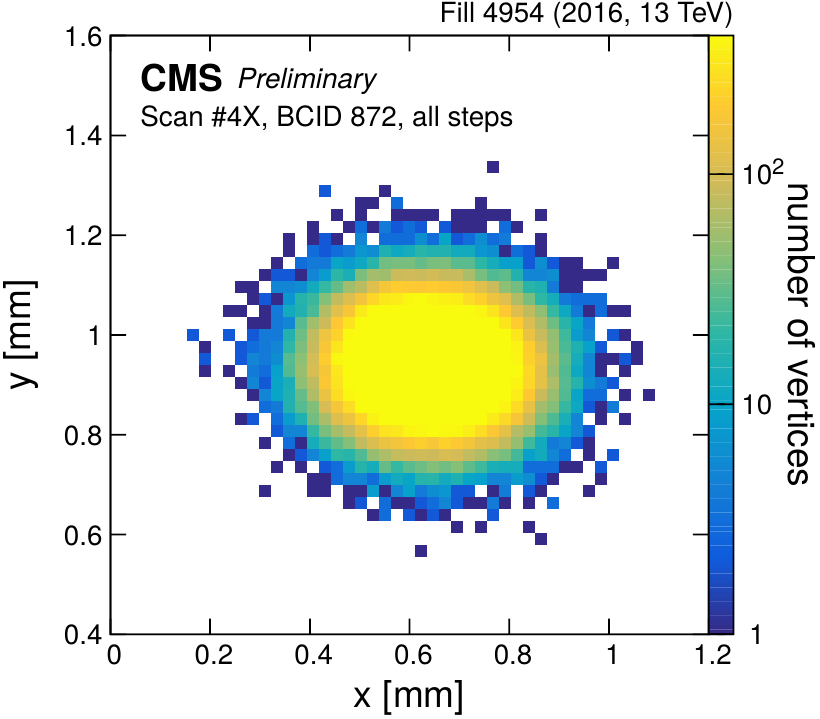}
\caption{Distribution of reconstructed transverse positions of the interaction vertices from all of the 19 scan steps of a beam-imaging scan.~\cite{DpsLumiDays}}
\label{fig:BeamImagingData}
\end{figure}

At each scan step, the CMS tracker system records data for reconstructing primary-interaction vertices~\cite{VertexSplitting}. Typically, a zero-bias trigger is used to store the data of five bunch crossings (identified by a number called ``BCID''), which are statistically independent and analyzed separately. Figure~\ref{fig:BeamImagingData} shows an example of the vertex distribution obtained from combining all steps of a beam-imaging scan.

\begin{figure*}[!t]
\centering
\includegraphics[width=\imagesize]{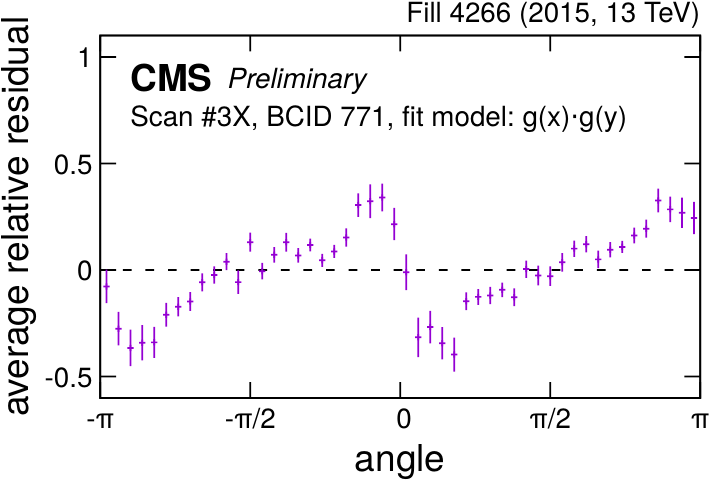}
\hspace*{\dimexpr\textwidth-2\columnwidth}
\includegraphics[width=\imagesize]{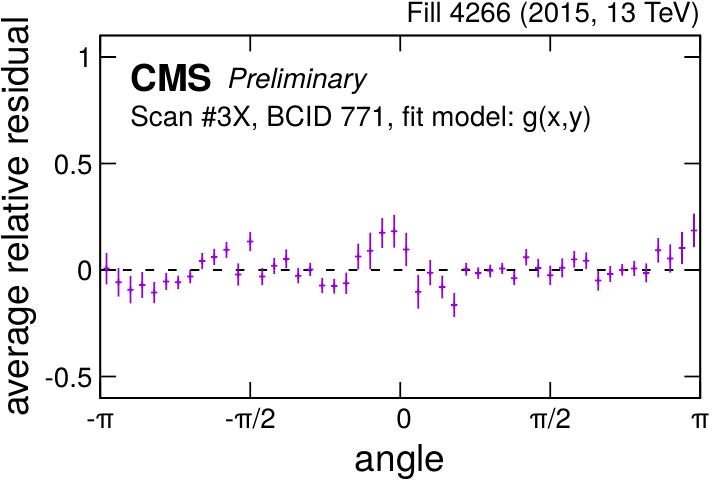}
\caption{Angular projection of the two-dimensional residual distribution of one of the beam images from a simultaneous fit using a single-Gaussian function as fit model with $\corr=0$ (left) and $\neq0$ (right).~\cite{DpsLumiDays}}
\label{fig:ResidualsAngular}
\end{figure*}

\begin{figure*}[!t]
\centering
\includegraphics[width=\imagesize]{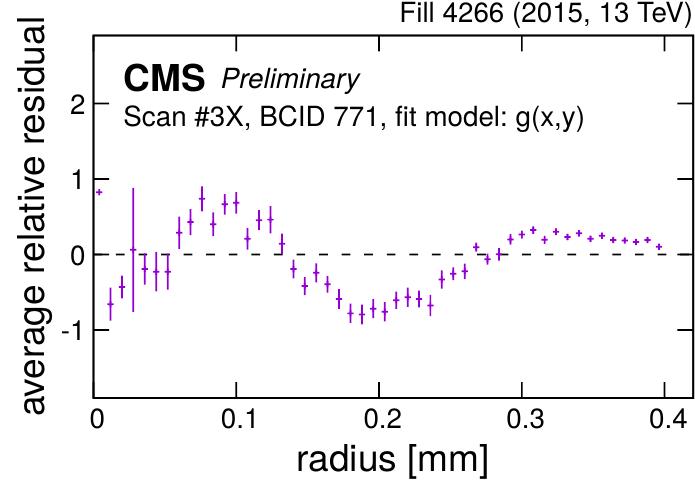}
\hspace*{\dimexpr\textwidth-2\columnwidth}
\includegraphics[width=\imagesize]{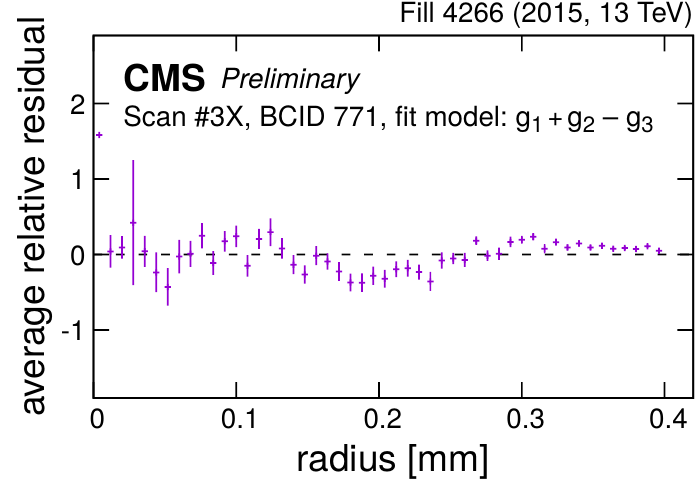}
\caption{Radial projection of the two-dimensional residual distribution of one of the beam images from a simultaneous fit using a single-Gaussian function (left) or a sum of the form $g_1+g_2-g_3$ (right) as fit model.~\cite{DpsLumiDays}}
\label{fig:ResidualsRadial}
\end{figure*}

The number of interaction vertices as a function of their transverse position, $\Nvtx(x,y)$, depends on the proton densities of the two LHC beams, $\rho_{1,2}$, and their transverse separation $\delx$:
\begin{align}
    \Nvtx(x,y;\delx)\propto\big[\rho_1(x+\delx,y)\cdot\rho_2(x,y)\big]\convV.
\end{align}
Here, ${}\convV$ denotes convolution with the vertex resolution function. For a small step size, the combination of vertex positions can be approximated as an integral over $\delx$:
\begin{align}
    \sum_\delx\Nvtx(x,y;\delx)&\approx\int_\delx\big[\rho_1(x+\delx,y)\cdot\rho_2(x,y)\big]\convV\,\text{d}\Delta x \nonumber\\
    &=\big[(\Mx\rho_1)(y)\cdot\rho_2(x,y)]\convV, \label{eq:beamimage}
\end{align}
where $\Mx\rho_1$ denotes the marginalization of the proton density along the scanning direction, here the $x$ coordinate. Thus, any nonfactorization present in $\rho_1$ is integrated out and Eq.~\eqref{eq:beamimage}, the ``beam image'', can be used to extract the nonfactorization of $\rho_2$.

\subsection{Fit Procedure}

The two beam proton densities $\rho_{1,2}$ are determined with a simultaneous fit to the four beam images of a beam-imaging scan set.

For the fit, the vertex resolution function $V$ is modeled with a two-dimensional Gaussian function. To allow for an analytical computation of the convolution~\cite{GaussianConvolution}, only fit models built from Gaussian functions are used to describe the beam proton densities. A two-dimensional Gaussian function has the form:
\begin{align}
    g(x,y)=w\exp\left[-\frac{1}{2(1-\corr^2)}\left(\frac{x^2}{\sigma_x^2}+\frac{y^2}{\sigma_y^2}-\frac{2\corr xy}{\sigma_x^{\vphantom{2}}\sigma_y^{\vphantom{2}}}\right)\right].
\end{align}
Here, $\sigma_{x,y}$ are the Gaussian widths along the two transverse coordinates, $\corr$ is the correlation parameter, and $w$ a weight. For $\corr\rightarrow0$, the Gaussian function factorizes into $x$- and $y$-dependent functions, $g(x,y;\corr=0)=g(x)\cdot g(y)$. For $\corr\neq0$, the factorization axis of the Gaussian function is rotated and thus a factorization bias is introduced for calibration results derived from VdM scans in the direction of $x$ and $y$.

Figure~\ref{fig:ResidualsAngular} shows example fit results using a single-Gaussian function as fit model, with either $\corr=0$ or $\neq0$. Typically, a small but nonzero value $0.01<|\corr|<0.15$ is required to describe the angular distribution of interaction vertices.

A higher nonfactorization occurs when the beam proton densities are described by sums of two or more Gaussian functions with different Gaussian widths. For the LHC fills~4266, 4954, and 6016 (calibration fills for proton-proton collisions at \sqrts{13} in 2015~\cite{LUM-15-001}, 2016~\cite{LUM-17-001}, and 2017~\cite{LUM-17-004}, respectively), the best description of the beam-imaging data is achieved by a sum of three Gaussian functions where the function with the smallest Gaussian width has a negative weight, $g_1+g_2-g_3$. Figure~\ref{fig:ResidualsRadial} shows example fit results using a single-Gaussian function and this model, where the latter gives a much better description of the radial distribution of interaction vertices.

The fit procedure is implemented using the \RooFit{} software~\cite{RooFit}. For proper convergence, the procedure requires positive definite model functions. While Gaussian functions and their sums are automatically positive definite, the sum $g_1+g_2-g_3$ with one negative weight requires a reparameterization to ensure definiteness. To further improve convergence, a two-step fit is employed: First, a double-Gaussian function $g_1+g_2$ is fitted. The resulting fit parameters are used as starting values for the positive components of the fit model $g_1+g_2-g_3$.

\subsection{Factorization Bias Evaluation}

The beam proton densities from the fit to the beam-imaging data are used in a Monte Carlo (MC) simulation of a VdM scan pair. Events with interaction vertices are generated for the 25 steps of both scans, reproducing the statistics and the setup of the actual VdM scans. A standard VdM analysis of this pseudodata is performed, yielding an estimate $\Ivdm$ of the beam overlap integral biased in the same way as the experimental VdM results.

A direct integration of the product of the two beam proton densities yields the unbiased beam overlap integral $\Itrue$. The difference between the two results, normalized to the unbiased value, $(\Ivdm-\Itrue)/\Itrue$, estimates the size of the factorization bias.

\begin{figure}[!h]
\centering
\includegraphics[width=\imagesize]{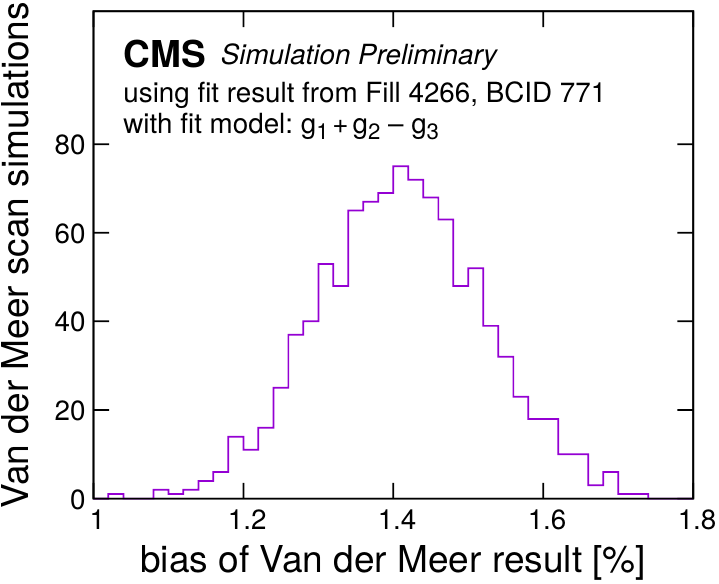}
\caption{Factorization bias of VdM results, evaluated for 1000 independent MC simulations of VdM scans, using the same set of parameters obtained from a fit to beam-imaging data. The standard deviation of this distribution is 0.11\,\%.~\cite{DpsLumiDays}}
\label{fig:VdmSimulation}
\end{figure}

Figure~\ref{fig:VdmSimulation} shows an example result of repeating this procedure 1000 times using the same set of beam-imaging fit results. The width of the distribution results from statistical fluctuations in the VdM scan data. It is typically $\approx$0.1\,\% and is taken as the uncertainty.

\subsection{Closure Test}

To evaluate the accuracy of the factorization bias estimate from the beam-imaging method, a closure test is performed. Toy models are constructed with randomly drawn parameter values. For each toy, events for a beam-imaging scan set are generated, and fits to this pseudodata are performed. The factorization bias is then obtained by a VdM simulation in the previously described way, either using the pseudodata fit results or the true toy model.

\begin{figure}[!ht]
\centering
\includegraphics[width=\imagesize]{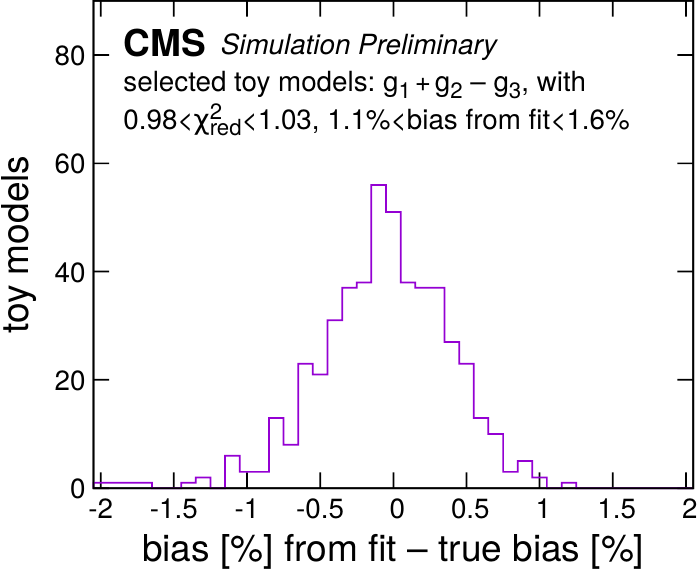}
\caption{Difference between bias obtained from direct integration of a toy model (``true bias''), and from a fit to pseudodata generated using the same toy model (``bias from fit''), for a selection of toys. The average difference is 0.06\,\%, with a standard deviation of 0.45\,\%.~\cite{DpsLumiDays}}
\label{fig:ClosureTest}
\end{figure}

Figure~\ref{fig:ClosureTest} shows the difference between both estimates, using a selection of toys whose fits to pseudodata have a similar obtained factorization bias and a similar reduced $\chi^2$ value as the fits to the five per-BCID datasets from LHC fill~4266. The closure test shows good agreement. The width of this distribution is taken as additional systematic uncertainty. For well-converging fits, it can be as low as 0.5\,\%, allowing for a precise determination of the factorization bias.

\subsection{Beam Position Effects}

The positions of the two beams during the VdM and beam-imaging scans are subject to several experimental effects. In the VdM method, the beam positions are corrected accordingly~\cite{LumiDaysMoritz}. For the beam-imaging method, the essential requirement is that one beam stays at a constant transverse position during all steps of one beam-imaging scan.

The length scale of the beam positions derived from the magnet steering are calibrated with beamspot measurements from the CMS tracker. This position scaling does not affect the essential beam-imaging requirement.

Orbit drift describes variations in the beam orbit during a nominally stable position~\cite{LumiDaysWitold}. The current orbit drift methodology does not allow us to assess the impact on the single beam's movement. However, some scenarios compatible with the observed orbit drift have been applied as corrections in the beam-imaging analysis. The difference to the bias obtained from the uncorrected fit results was found to be small compared to its systematic uncertainty.

Beam-beam deflection refers to the electric repulsion of the beams, which changes the position of the nominally non-moving beam depending on the transverse beam separation. The size of this effect is predicted analytically~\cite{LumiDaysVladik,LumiDaysTatiana}. The largest differences between the non-moving beam positions are typically smaller than 4\,\mum, much smaller than the typical step size of the moving beam of about 50\,\mum. An ongoing study evaluates the impact of the beam-beam deflection on the estimate of the factorization bias.

\section{Offset Scan Analysis}

An offset scan is a VdM scan variant where the two beams are additionally kept at constant separation in the non-scanning transverse direction. Since VdM and offset scans test different transverse slices of the beam overlap region, they are affected differently by nonfactorization.

\begin{figure}[!h]
\centering
\includegraphics[width=\imagesize]{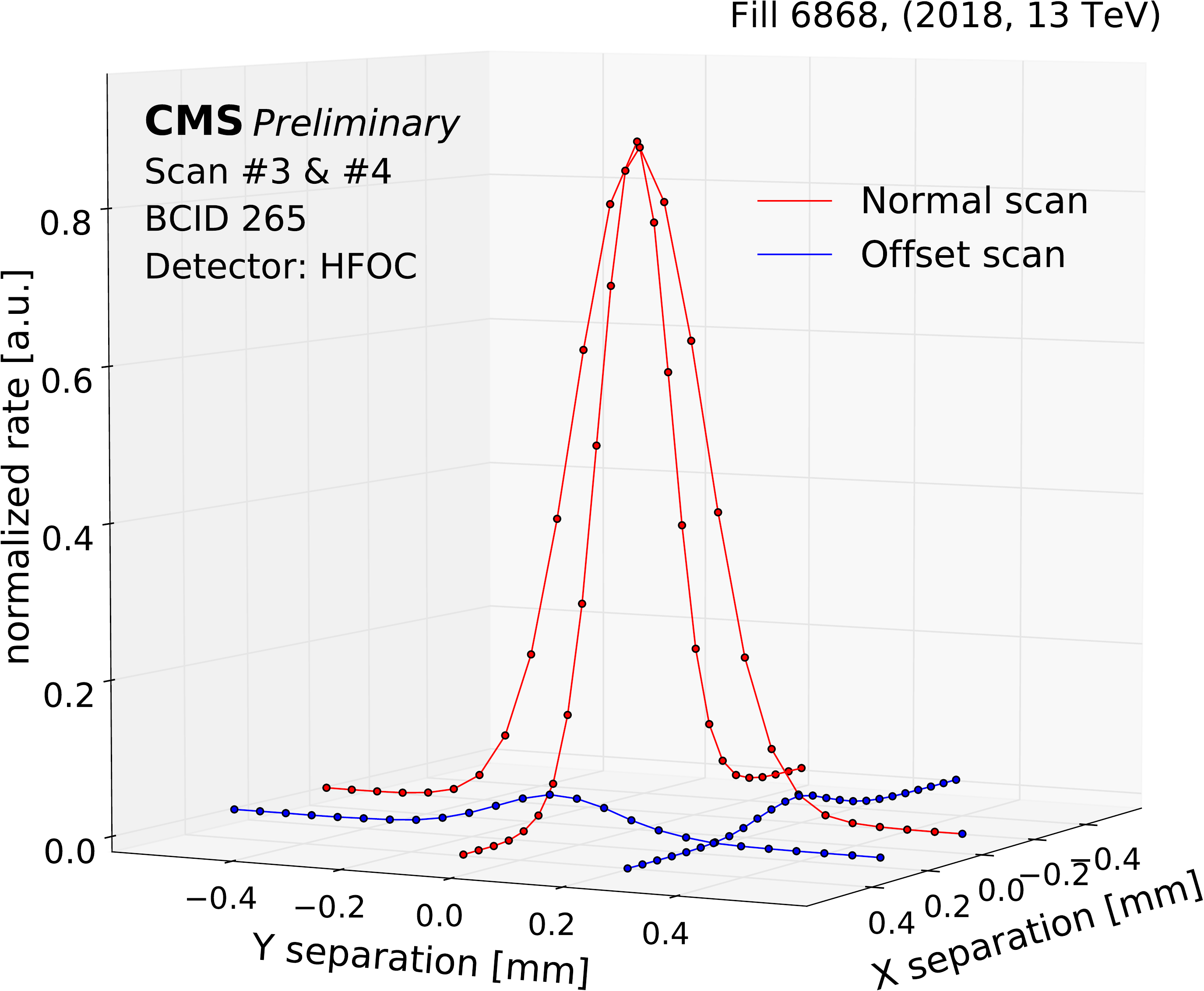}
\caption{Interaction rate as function of transverse beam separations for an example of VdM and offset scan pairs.~\cite{DpsLumiDays}}
\label{fig:OffsetScans}
\end{figure}

The offset scan analysis~\cite{AtlasLum8Tev} performs a simultaneous fit to the rates measured as a function of the transverse beam separations during VdM and offset scan pairs, reconstructing the transverse profile of the beam overlap region. Figure~\ref{fig:OffsetScans} shows an example of the collected data.

To correct for the beam position effects in the offset scan analysis, a position correction is introduced as a free parameter of the fit. This parameter equalizes the rates measured for steps of different scans with the nominally same beam separation in both transverse directions.

Different models can be used for the transverse profile of the beam overlap region. For the LHC fill~6868 (calibration fill for proton-proton collisions at \sqrts{13} in 2018~\cite{LUM-18-002}), a good description of the data is achieved with single- and double-Gaussian functions. The output of the fit procedure is the correlation parameter $\corr$ of the main Gaussian function. Compared to preliminary fit results of the beam-imaging scans in the same fill, a good agreement in the extracted value of $\corr$ is observed. Independent analyses of scans performed 12 hours apart during that fill show that the value of $\corr$ increased over time.

Testing additional components of the transverse proton densities of the LHC beams, as suggested by the results of the beam-imaging method, is limited by the low event count in large beam separations of the offset scans. Further studies need to evaluate whether the offset scan analysis is capable of extracting the full nonfactorization information.

\section{Beamspot Evolution Analysis}

Complementary to the beam-imaging method, the analysis of the beamspot evolution~\cite{BeamspotEvolution} aims at reconstructing the three-dimensional proton densities of the two LHC beams. It exploits a three-dimensional beamspot fit~\cite{BeamspotFits} of the interaction vertices separately for each scan step of a VdM scan pair, extracting the mean position and the width of the distribution in each of the three spatial coordinates, and the tilts of the distribution between any two spatial coordinates. These observables are fitted together with the measured event rates as a function of the transverse beam separation, extracting the beam proton densities.

The beamspot evolution analysis has been applied by the ATLAS and ALICE experiments to estimate the size of the factorization bias~\cite{LumiDaysMateusz}. In an ongoing study, the beamspot evolution analysis is applied to VdM scan data recorded by the CMS experiment, and its outcome is used to cross-check the beam-imaging method.

\section{Diagonal Scan Analysis}

For the diagonal scan analysis, four VdM scans are performed along four transverse directions rotated with respect to each other by 45\textdegree. Thus, in addition to the standard VdM scans along $x$ and $y$, additional diagonal VdM scans are performed along $x+y$ and $x-y$.

\begin{figure}[!h]
    \centering
    \includegraphics[width=\imagesize]{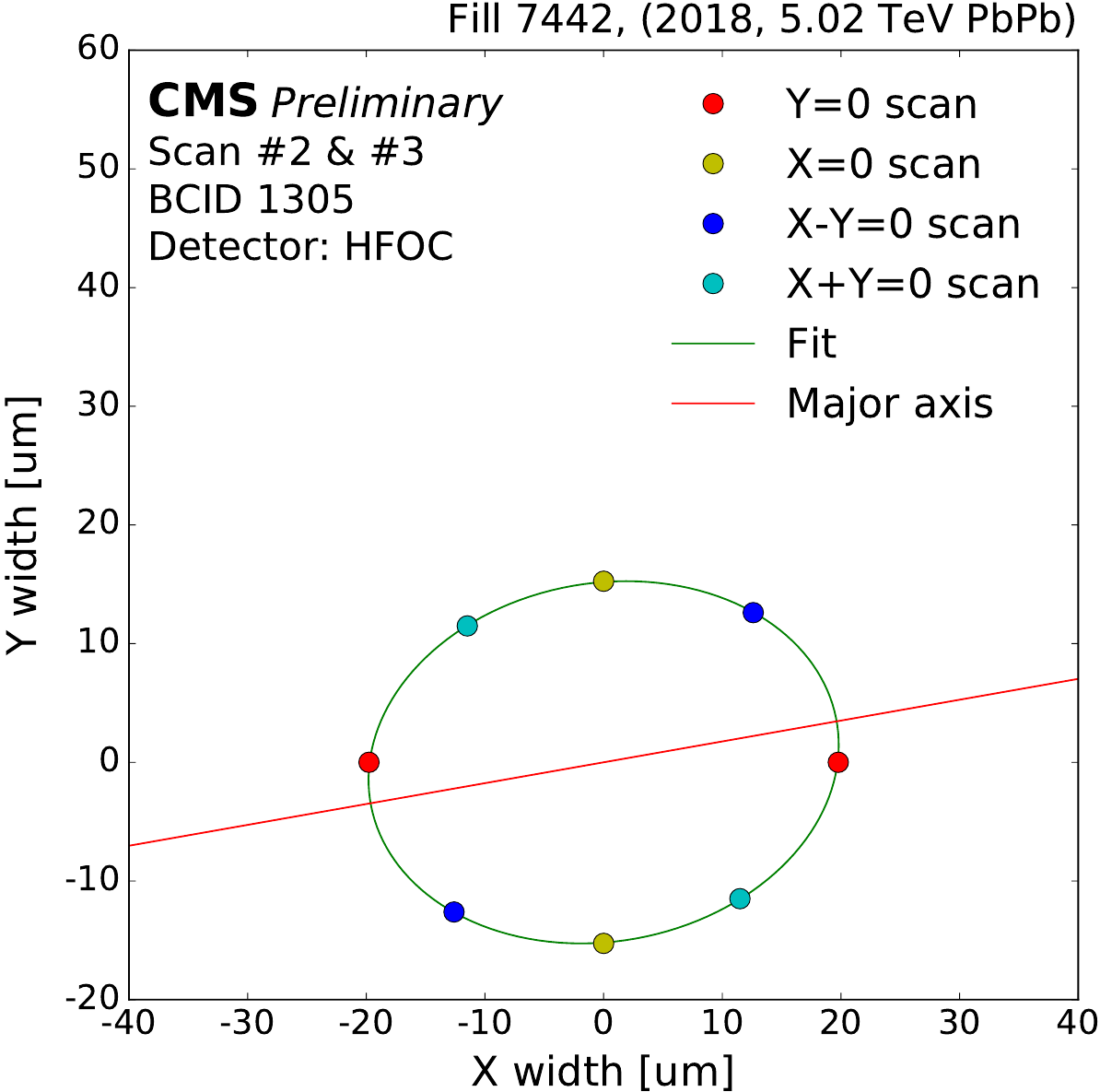}
    \caption{Beam overlap region width along different scan directions for a set of VdM and diagonal scans. Points with same color belong to the same scan.~\cite{DpsLumiDays}}
    \label{fig:DiagonalScans}
\end{figure}

The width of the beam overlap region is determined separately for each of the four scan directions in the standard way. Figure~\ref{fig:DiagonalScans} shows fit results for the LHC fill~7442 (calibration fill for lead-lead collisions at \sqrts[\text{NN}]{5.02} in 2018). The four data points are fitted with an ellipse, yielding the one standard-deviation contour of the beam overlap region. As a result, a correlation parameter $\corr$ can be extracted as an estimate for the strength of the nonfactorization.

Diagonal scans have been performed for the first time at the CMS experiment in this data-taking period because the event rates in lead-lead collisions would be too low to get useful data from offset scans and the beam width of the lead beams is too small for the beam imaging method.

\section{Summary}

\begin{figure}[!h]
\includegraphics[width=\imagesize]{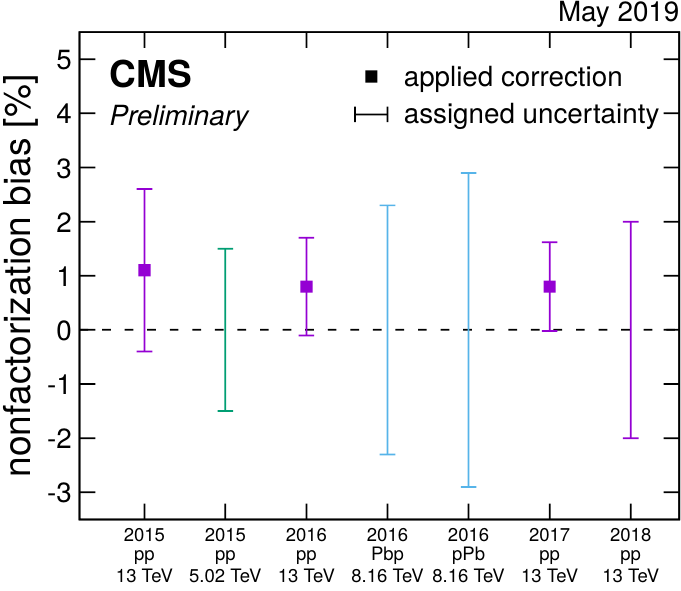}
\caption{Factorization bias corrections applied and systematic uncertainties assigned to the visible cross sections measured in VdM scans for the different \Run2 data-taking periods. In some cases, no correction is applied. \cite{DpsLumiDays}}
\label{fig:Summary}
\end{figure}

The methods applied by the CMS experiment to evaluate and correct for the factorization bias of the calibration of the luminosity measurement with Van der Meer scans have been presented. The preliminary results for the different data-taking periods in \Run2 are summarized in Fig.~\ref{fig:Summary}, and rely on the beam-imaging method.

Ongoing studies on the beam-imaging method suggest that updated results with a precision of 0.5\,\% are feasible for some of the data-taking periods.

Additionally, the CMS experiment exploits the analysis of offset scans, luminous region (``beamspot'') evolution, and diagonal scans. This will allow to independently cross-check the beam-imaging method.


\begin{thebibliography}{99}
\newcommand\Author[1]{#1,}
\newcommand\Title[1]{``#1'',}
\newcommand\Journal[3]{\textit{#1} \textbf{#2} #3,}
\newcommand\OtherReport[2]{#1, Rep. #2,}
\newcommand\Report[1]{\OtherReport{CERN, Geneva, Switzerland}{#1}}
\newcommand\Proceedings[2]{in \textit{Proc. #1}, #2,}
\newcommand\PresentedHere{presented at the LHC Lumi Days 2019, Prévessin, France, June 2019, this conference.}
\newcommand\Date[1]{#1.}
\newcommand\Doi[1]{\href{http://doi.org/#1}{\texttt{doi:#1}}}
\newcommand\Cds[1]{\href{http://cds.cern.ch/record/#1}{\texttt{cds.cern.ch/record/#1}}}
\newcommand\Arxiv[2]{\href{http://arxiv.org/abs/#1}{\texttt{arXiv:#1 [#2]}}}
\newcommand\Manchester{Manchester U., Manchester, UK}

\bibitem{CMS-Experiment}
    \Author{CMS Collaboration}
    \Title{The CMS experiment at the CERN LHC}
    \Journal{JINST}{3}{S08004}
    \Date{Aug. 2008}
    \Doi{10.1088/1748-0221/3/08/S08004}

\bibitem{LumiDaysMoritz}
    \Author{M. Guthoff (CMS Collaboration)}
    \Title{Overview of the CMS \Run2 luminosity determination and calibration methodology}
    \PresentedHere

\bibitem{LUM-13-001}
	\Author{CMS Collaboration}
    \Title{CMS luminosity based on pixel cluster counting -- summer 2013 update}
    \Report{CMS-PAS-LUM-13-001}
    \Date{Sep. 2013}
    \Cds{1598864}

\bibitem{LhcbLum8Tev}
    \Author{LHCb Collaboration}
    \Title{Precision luminosity measurements at LHCb}
    \Journal{JINST}{9}{P12005}
    \Date{Dec. 2014}
    \Doi{10.1088/1748-0221/9/12/P12005}

\bibitem{AtlasLum8Tev}
    \Author{ATLAS Collaboration}
    \Title{Luminosity determination in pp collisions at \sqrts{8} using the ATLAS detector at the LHC}
    \Journal{Eur. Phys. J. C}{76}{653}
    \Date{Nov. 2016}
    \Doi{10.1140/epjc/s10052-016-4466-1}

\bibitem{AliceLum8Tev}
	\Author{ALICE Collaboration}
    \Title{ALICE luminosity determination for pp collisions at \sqrts{8}}
    \Report{ALICE-PUBLIC-2017-002}
    \Date{Mar. 2017}
    \Cds{2255216}

\bibitem{BeamImaging}
	\Author{M.~Klute, C.~Medlock, and J.~Salfeld-Nebgen}
    \Title{Beam imaging and luminosity calibration}
    \Journal{JINST}{12}{P03018}
    \Date{Mar. 2017}
    \Doi{10.1088/1748-0221/12/03/P03018}

\bibitem{VertexSplitting}
	\Author{CMS Collaboration}
    \Title{Description and performance of track and primary-vertex reconstruction with the CMS tracker}
    \Journal{JINST}{9}{P10009}
    \Date{Oct. 2014}
    \Doi{10.1088/1748-0221/9/10/P10009}

\bibitem{DpsLumiDays}
    \Author{CMS Collaboration}
    \Title{Nonfactorization in Van der Meer scans in \Run2}
    \Report{CMS-DP-2019-019}
    \Date{Jul. 2019}
    \Cds{2681801}

\bibitem{GaussianConvolution}
    \Author{P. Bromiley}
    \Title{Products and convolutions of Gaussian probability density functions}
    \OtherReport{\Manchester}{TINA Memo 2003-003}
    \Date{Jun. 2018}
    \href{http://tina-vision.net/docs/memos/2003-003.pdf}{\texttt{tina-vision.net/docs/memos/2003-003.pdf}}

\bibitem{LUM-15-001}
	\Author{CMS Collaboration}
    \Title{CMS luminosity measurement for the 2015 data-taking period}
    \Report{CMS-PAS-LUM-15-001}
    \Date{Mar. 2016}
    \Cds{2138682}

\bibitem{LUM-17-001}
	\Author{CMS Collaboration}
    \Title{CMS luminosity measurements for the 2016 data-taking period}
    \Report{CMS-PAS-LUM-17-001}
    \Date{Mar. 2017}
    \Cds{2257069}

\bibitem{LUM-17-004}
	\Author{CMS Collaboration}
    \Title{CMS luminosity measurement for the 2017 data-taking period at \sqrts{13}}
    \Report{CMS-PAS-LUM-17-004}
    \Date{Jun. 2018}
    \Cds{2621960}

\bibitem{RooFit}
    \Author{W. Verkerke and D. Kirkby}
    \Title{The \RooFit{} toolkit for data modeling}
    \Proceedings{Computing in High Energy and Nuclear Physics (CHEP03)}{La Jolla, CA, USA}
    \Date{Mar. 2003}
    \Arxiv{physics/0306116}{physics.data-an}

\bibitem{LumiDaysWitold}
    \Author{W. Kozanecki}
    \Title{Impact of orbit perturbations on luminosity calibrations}
    \PresentedHere

\bibitem{LumiDaysVladik}
    \Author{V. Balagura}
    \Title{Beam-beam correction in VdM scans}
    \PresentedHere

\bibitem{LumiDaysTatiana}
    \Author{T. Pieloni and C. Tambasco}
    \Title{Beam-beam effects in Van der Meer scans: COMBI and TRAIN beam-beam corrections}
    \PresentedHere

\bibitem{LUM-18-002}
	\Author{CMS Collaboration}
    \Title{CMS luminosity measurement for the 2018 data-taking period at \sqrts{13}}
    \Report{CMS-PAS-LUM-18-002}
    \Date{May 2019}
    \Cds{2676164}

\bibitem{BeamspotEvolution}
	\Author{S. Webb}
    \Title{Factorization of beams in Van der Meer scans and measurements of the $\smash{\phi^\ast_\eta}$ distribution of $Z\rightarrow e^+e^-$ events in pp collisions at \sqrts{8} with the ATLAS detector}
    \OtherReport{Ph.D. thesis, \Manchester}{CERN-THESIS-2015-054}
    \Date{May 2015}
    \Cds{2020875}

\bibitem{BeamspotFits}
	\Author{CMS Collaboration}
    \Title{Beamspot measurements in 2016}
    \Report{CMS-DP-2016-051}
    \Date{Sep. 2017}
    \Cds{2285428}

\bibitem{LumiDaysMateusz}
    \Author{M. Dyndal (ATLAS and ALICE Collaborations)}
    \Title{Non-factorization in ATLAS \& ALICE VdM scans}
    \PresentedHere

\end{thebibliography}
\end{document}